\documentclass[journal=jpclcd,manuscript=letter,layout=twocolumn]{achemso}
\usepackage{chemformula}
\usepackage[T1]{fontenc}
\usepackage{amsmath}
\usepackage{amssymb}

\usepackage{epsfig,bm}

\author{Francesco Cordero}
\email{francesco.cordero@ism.cnr.it}
\author{Floriana Craciun}
\affiliation[ARTOV]
{Istituto di Struttura della Materia-CNR (ISM-CNR), Area della Ricerca di Roma - Tor Vergata, Via del Fosso del Cavaliere 100, I-00133 Roma, Italy}
\author{Francesco Trequattrini}
\affiliation[Fis]
{Dipartimento di Fisica, Universit\`{a} di Roma "La Sapienza", p.le A. Moro 2, I-00185 Roma, Italy}
\alsoaffiliation[ARTOV]
{Istituto di Struttura della Materia-CNR (ISM-CNR), Area della Ricerca di Roma - Tor Vergata, Via del Fosso del Cavaliere 100, I-00133 Roma, Italy}
\author{Amanda Generosi}
\author{Barbara Paci}
\affiliation[ARTOV]
{Istituto di Struttura della Materia-CNR (ISM-CNR), Area della Ricerca di Roma - Tor Vergata, Via del Fosso del Cavaliere 100, I-00133 Roma, Italy}
\author{Anna Maria Paoletti}
\author{Giovanna Pennesi}
\affiliation[MLIB]
{Istituto di Struttura della Materia-CNR (ISM-CNR), Area della Ricerca di Roma 1, Via Salaria, Km 29.300, I-00015 Monterotondo Scalo, Roma, Italy}

\title[Stability of Cubic FAPbI$_3$]{Stability of Cubic FAPbI$_3$ from X-Ray
Diffraction, Anelastic and Dielectric Measurements}

\begin{document}

\begin{tocentry}
\includegraphics[width=5 cm]{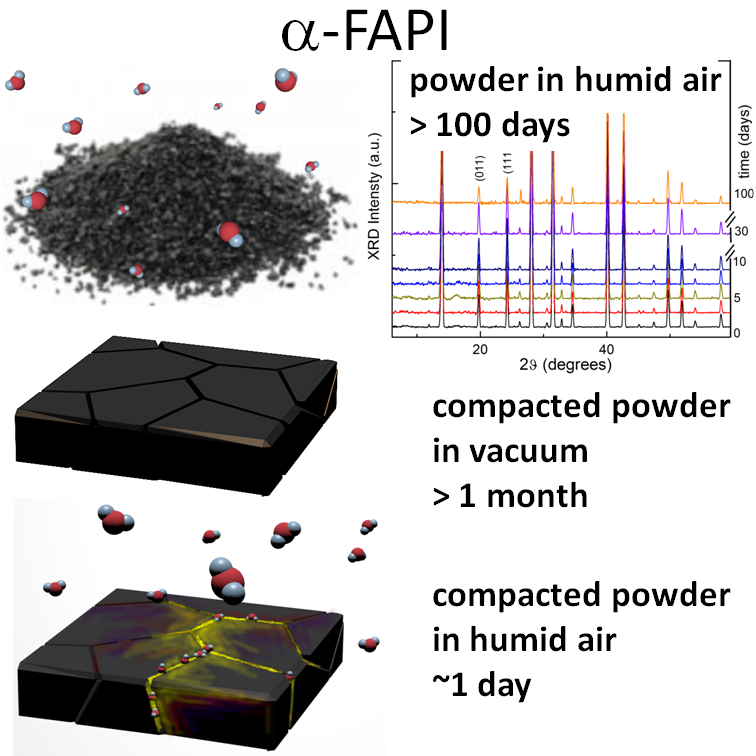}
\end{tocentry}

\begin{abstract}
Among the hybrid metal-organic perovskites for photovoltaic applications
FAPbI$_{3}$ (FAPI) has the best performance regarding efficiency and the
worst regarding stability, even though the reports on its stability are
highly contradictory. In particular, since at room temperature the cubic $%
\alpha $ phase, black and with high photovoltaic efficiency, is metastable
against the yellow hexagonal $\delta $ phase, it is believed that $\alpha -$%
FAPI spontaneously transform into $\delta -$FAPI within a relatively short
time. We performed X-ray diffraction and thermogravimetric measurements
on loose powder of FAPI, and present the first complete dielectric
and anelastic spectra of compacted FAPI samples under
various conditions. We found that $\alpha -$FAPI is perfectly stable for at
least 100 days, the duration of the experiments, unless extrinsic factors
induce its degradation. In our tests, degradation was detected after
exposure to humidity, strongly accelerated by grain boundaries and the
presence of $\delta $ phase, but it was not noticeable on the loose powder
kept in air under normal laboratory illumination. These findings have strong
implications on the strategies for improving the stability of FAPI without
diminishing its photovoltaic efficiency through modifications of its
composition.
\end{abstract}



Although MAPbI$_{3}$ (MAPI, MA = methylammonium CH$_{3}$NH$_{3}$) is the
most studied hybrid metal-organic perovskite for photovoltaic applications,%
\cite{KPP18}
better performance in terms of photovoltaic efficiency are found in FAPbI$%
_{3}$ (FAPI, FA = formamidinium CH(NH$_{2}$)$_{2}$). This is due both to a
smaller bandgap of FAPI and to the fact that the FA$^{+}$ ion, in spite of a
smaller electric dipole with respect to MA$^{+}$, has a much larger
quadrupole and faster reorientation dynamics that better screen the
photoexcited carriers, enhancing their lifetime.\cite{CCF17c} FAPI also has
a better stability than MAPI at high temperature but its major flaw is that
the black cubic $\alpha $ phase, which has the high photovoltaic efficiency,
is metastable at room temperature, where instead the stable phase, and the
one obtained by standard chemical methods, is the yellow hexagonal $\delta $
phase. For these reasons, major efforts are directed now at trying to
stabilize the cubic $\alpha $ phase of FAPI through partial substitutions of
FA with MA, Cs, etc. or I with Br, although this approach increases the
bandgap.\cite{BHD15,FWW17} It has been discussed, based on neutron
diffraction measurements and simulations, that the $\alpha \rightarrow
\delta $ transformation is complex and occurs through various intermediate
stages, requiring to overcome a free energy barrier estimated in the order
of hundreds of milli--electron volt.\cite{CFP16} This explains why the $%
\alpha $ phase of FAPI is kinetically trapped, resulting in a large thermal
hysteresis between the $\delta \rightarrow \alpha $ transition at $T_{\delta
\alpha }^{h}=350$~K and the $\alpha \rightarrow \delta $ at $T_{\delta
\alpha }^{c}=290$~K.\cite{CFP16} Actually, the barrier for the $\alpha
\rightarrow \delta $ transition has not been measured, and there is complete
uncertainty on the kinetics of this transition. Indeed, also the reported
temperatures for the $\delta \rightarrow \alpha $ transition during heating
span quite a broad range between 350~K\cite{CFP16} and 458~K,\cite{HBS16}
and this may be partly explained with the slow kinetics of the
transformation, since the lowest temperature is observed in a quasistatic
neutron scattering experiment while the highest during fast heating at
5~K/min. The uncertainties on the reverse transformation are worse. The
reported stability of the $\alpha $ phase at room temperature in humid or
dessicated air ranges from few hours,\cite{LKK15,ZKG16,LYP16,ZWJ16,ZZY18} to
few days,\cite{JNY15,SND16,GZK18} but stability over months has been
achieved by growing FAPI in a solution with long chain alkyl or aromatic
ammonium cations which cover the $\left( 100\right) $ grain surfaces and
render them stable,\cite{FWW17} or by growing FAPI within template alumina
nanotubes.\cite{GZK18} At variance with studies that highlight the role of
defects and grain boundaries in stimulating degradation,\cite{AWH16,YKY18}
it is also believed that nanostructured FAPI has a superior
stability.\cite{LOT17,ZCQ18} Closely connected with this issue is the variety of
results obtained in different experiments using a technique as effective as
neutron diffraction for studying the nature of the polymorphism of FAPI, as
recently summarized in Ref. \citenum{WGG18}.

We will show that a loose micrometric powder of FAPI, prepared by slightly
modifying the known procedures can remain exposed in air at normal
laboratory illumination for at least 100 days without any sign of
degradation and pressed pellets can likewise remain in the $\alpha $ phase
for at least one month after being thoroughly dried in high vacuum. This
observations suggest that the kinetic trapping of the $\alpha $ phase of
FAPI at room temperature lasts for much longer than previously thought, and
the transformation to the $\delta $ phase, itself highly hygroscopic, is
catalyzed by humidity together with other extrinsic factors, such as grain
boundaries.


FAPI samples were obtained following an already reported procedure\cite%
{LSY16} with slight modifications, by precipitating with toluene the yellow phase
from a solution of HC(NH$_2$) I$_2$ and PbI$_2$ in GBL. The yellow powder
was recovered by centrifugation and put under vacuum at $150\div 160$~${}^{\circ }$C
for $\sim 4$~h until completely transformed into dark grey/black $\alpha-$FAPI
(details in SI).

We performed four types of experiments on two types of samples: X-ray
diffraction (XRD) in air at room temperature on loose powder and also on the
compacted samples; complex Young's modulus versus temperature in high vacuum
on a compacted thin bar (FAPI-b) and dielectric permittivity versus
temperature in a small closed volume on compacted discs (FAPI-d1 and
FAPI-d2). The results on the stability of $\alpha -$FAPI turned out to be
strongly dependent on the different experimental conditions.


In order to test whether traces of water or other solvents were present in
the freshly prepared powders, we performed Thermo Gravimetric Analysis (TGA)
and Differential Scanning Calorimetry (DSC) on $20-30$~mg batches of $\delta
-$FAPI-before and after its conversion to $\alpha -$FAPI. The results are
shown in Fig. S1 and no mass loss attributable to water or other solvents
could be detected until the sample decomposition started above 250~${{}^\circ%
}$C. The yellow samples exhibited a dip in the DSC signal at $\gtrsim 150$~${%
{}^\circ}$C, due to the $\delta \rightarrow \alpha $ conversion.

\begin{figure}[tbh]
\includegraphics[width=8.5 cm]{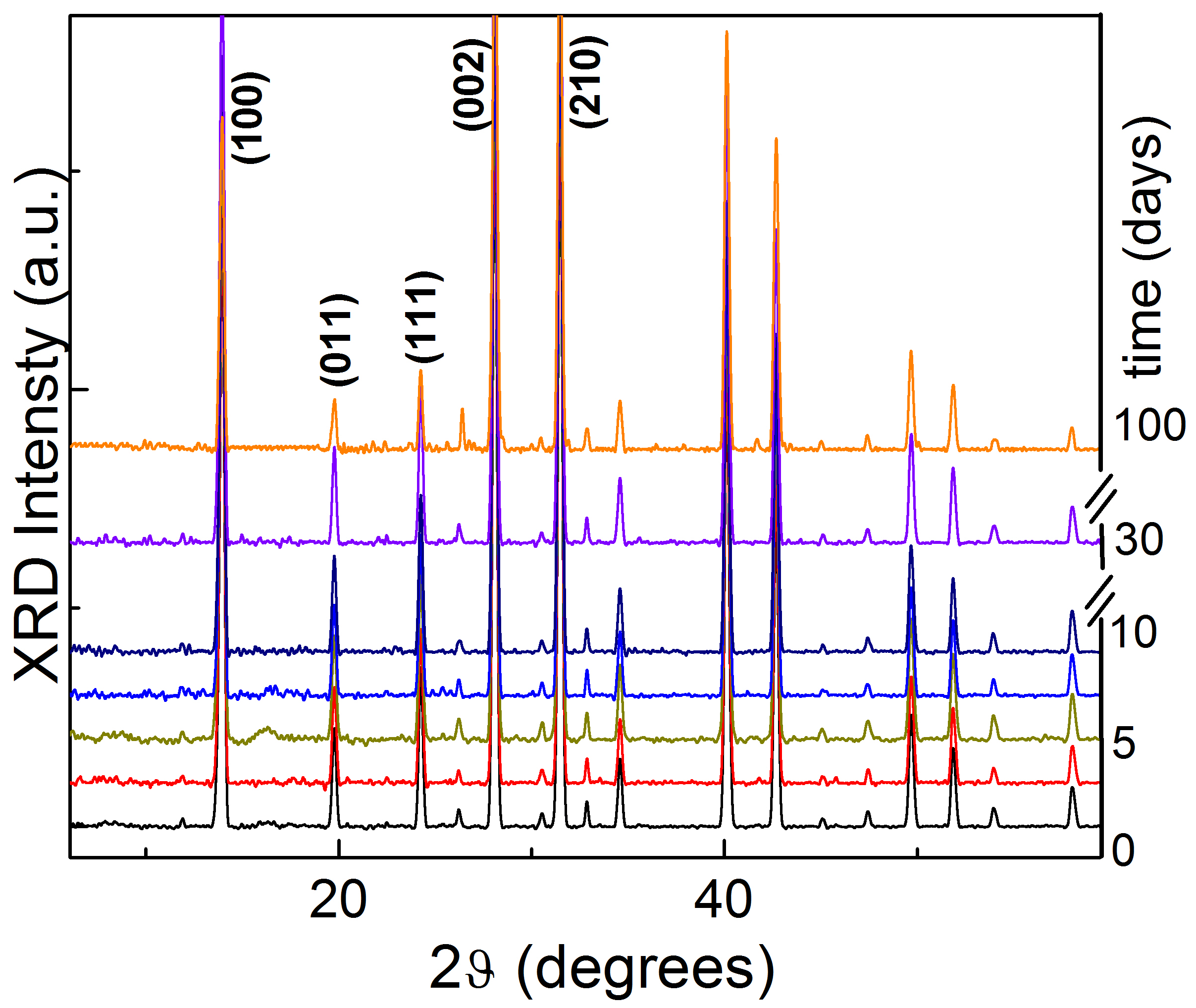}
\caption{XRD patterns collected from the precursor FAPI powder at different
times. The $\protect\alpha -$FAPI cubic phase was detected and labeled
according to the literature.\protect\cite{WWF15,XCN17}}
\label{fig-XRDp}
\end{figure}

An unexpected positive result was the perfect stability for at least 97 days of
the loose powder of $\alpha $-FAPI in air under normal laboratory
illumination with neon lamps. In Fig. \ref{fig-XRDp} the XRD spectra taken
over a period of 100 days did not change at all (during the 2 months
before the last measurement the powder was kept in a closed polypropylene
vial). The initial pattern was acquired few hours after the conversion of
the $\delta -$FAPI powder into $\alpha -$FAPI. The particles
had sizes exceeding the micrometer, as deduced from the narrow reflections,
whose width was determined by the XRD setup (SI). This is plain evidence that the
kinetics of the $\delta \rightarrow \alpha $ transition in bulk $\alpha -$%
FAPI at room temperature is much slower than generally assumed, even without
the often cited stabilization size effect in nanometric particles.\cite%
{LOT17,ZCQ18} This is true not only above $T_{\delta \alpha }^{c}\simeq 290$%
~K deduced from neutron diffraction, \cite{CFP16} but also including one
week at several degrees below that temperature. The stability against
decomposition is likewise high, since no trace of PbI$_{2}$ is detected.
The first sign of degradation through reversion to the $\delta$ phase has
been found after additional 45 days (Fig. S2).

On the other hand, the small part of powder that remained attached to
the walls of the glass bottle rapidly became yellow (SI).


The pressures at which we compacted the discs and the bars were 0.62 and
0.2~GPa respectively, in the range of structural transitions of $\alpha $%
-FAPI reported in the literature, notably to the $\delta $ phase,\cite{JLJ18}
even though there is disagreement.\cite{WGG17,LKG17} Indeed, the presence of
$\delta $-FAPI in the pressed samples could be determined both visually, as
yellow spots (see Fig. S3), and from the XRD analysis of the disc FAPI-d2
(Fig. S2).
It seems that also grinding the powder of originally pure $\alpha $-FAPI in
air rapidly induced a transformation to $\delta $-FAPI. In fact, being the
first attempts to make a bar unsuccessful, we repeatedly pressed and
reground the powder in an agate mortar, observing a progressive enrichment
in yellow grains. The bar finally used in the anelastic experiments (FAPI-b)
was pressed from freshly prepared powder of $\alpha -$FAPI in a single
successful attempt.

\begin{figure}[tbh]
\includegraphics[width=8.5 cm]{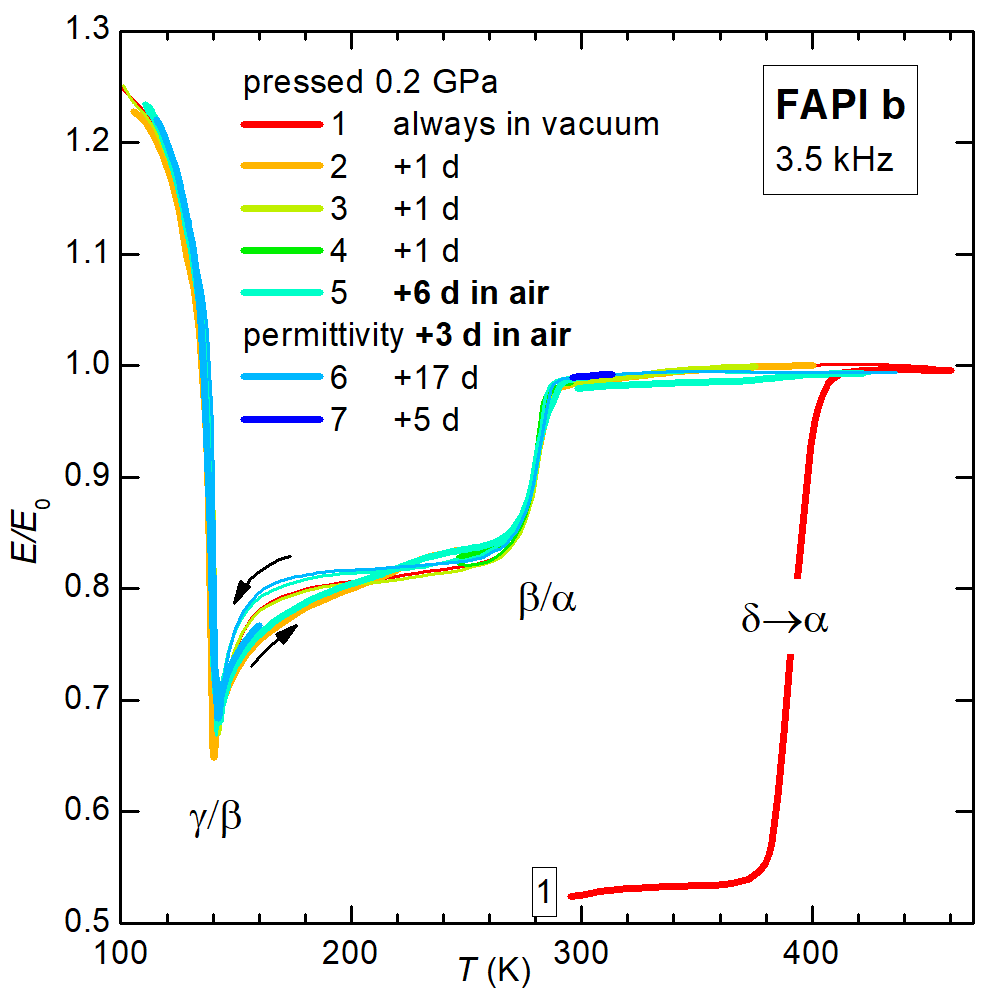}
\caption{Young's modulus, normalized to the initial value in the $\protect%
\alpha $ phase, measured at 3.5~kHz over several temperature cycles during
35 days. The sample was always in vacuum, except for 6 days between curves 4
and 5 and additional 3 days for measuring the dielectric permittivity
before curve 6. See SI for details.}
\label{fig-anel}
\end{figure}

The anelastic measurements were performed in a vacuum of $<10^{-5}$~mbar at
high temperature or in $<0.2$ mbar He at low temperature, and the system was
left in static vacuum between the experiments. Details on the technique and
sample conditions are found in SI.
The anelastic spectrum of FAPI has been measured here for the first time,
similarly to the complete dielectric spectrum, and both are used here for the
specific task of evaluating the fraction of $\alpha$ phase in our samples.
The conclusions that can be drawn from these spectra on the nature of the
structural transitions that $\alpha-$FAPI undergoes are left for future
work.

Figure \ref{fig-anel} presents most of these measurements as Young's modulus
$E$ normalized to the value $E_{0}$ initially reached in the $\alpha $
phase. During the first heating, a sudden rise of the modulus at $\sim 390$%
~K must correspond to the transformation of the $\delta $ phase induced by
pressure back to the $\alpha $ phase, after which the sample remains in the $%
\alpha $ phase for all the subsequent temperature cycles and agings at room
temperature.
When cooling, the $\alpha $ phase transforms into $\beta $ at
280~K\cite{FSL16} and then into $\gamma $ at 141~K.\cite{FSL16} The
anomalies occurring at these temperatures are retraced during heating,
except for some thermal hysteresis above the $\gamma \rightarrow \beta $
transition, as indicated by the arrows. It is not the aim of this paper to
analyze the elastic anomalies at these phase transitions; we will only say
that the steplike softening at the $\alpha \rightarrow \beta $ transition is
as expected for a transition where the PbI$_{6}$ octahedra tilt with respect
to the symmetric cubic phase (strain is coupled to the square of the tilt
angle\cite{Reh73}), and is analogous to what happens in MAPI.\cite{CCT18}
The transition to the $\gamma $ phase is more enigmatic, defined as
reentrant psudocubic\cite{FSL16} with near cancelation of the octahedral
tilting and presumably freezing of the reorientations of the FA molecules.
Such a freezing would explain the restiffening of the elastic modulus well above the
value of the cubic phase, and is similar to what happens below the analogous
transition in MAPI, although in that case the reorientational freezing
is mainly thermally activated and occurs below the transition to the
$\gamma $ phase.

The interesting point in the present context is that, after the initial $%
\delta \rightarrow \alpha $ transformation, the $E\left( T\right) $ curves
remain perfectly stable for 1 month, and this excludes that any transformation to
the $\delta $ phase or degradation occurs over the month time scale in
vacuum.
A minor degradation, indicated by $\lesssim 1\%,$ decrease of $E$, occurred
during a total of 9 days in air between curves 4 and 6, during which
the dielectric permittivity was also measured. Details are found in the SI.

Few days after completing the cycle of anelastic measurements, the XRD
spectra showed that a rapid degradation processes had started. The $\alpha $
phase was still present but the dominant peak was (001) of PbI$_{2}$ (Fig.
S2). The high degree and irreversibility of degradation was confirmed by a
new cycle of anelastic measurements (Fig. S6) with an extremely low Young's
modulus that could not be recovered by annealing above 400~K.

\begin{figure}[tbh]
\includegraphics[width=8.5 cm]{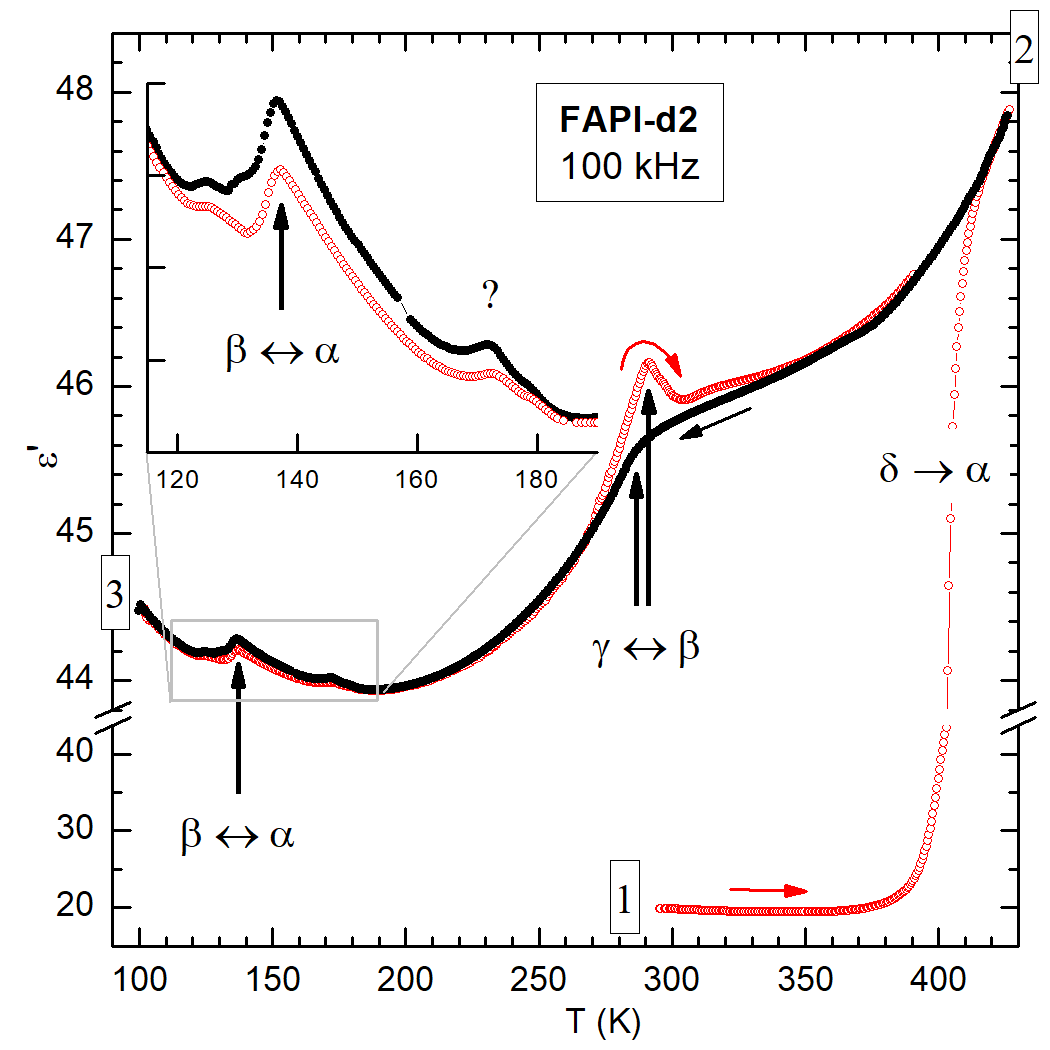}
\caption{Real part of the permittivity of FAPI measured at 100~kHz during
(1) heating at 1.2~K/min 2~h after pressing the disc, (2)\ cooling at $-1.2$%
~K/min and (3) heating at 1.5~K/min. Note the broken vertical scale.}
\label{fig-diel3}
\end{figure}

The complex dielectric permittivity was measured in a Linkam stage with
modified Examina Probe having a sealed volume of $\sim 100$~cm$^{3}$. The
standard procedure for removing humidity is flushing the probe with a dry
gas evaporated from liquid nitrogen or air, for few minutes at 40~${{}^\circ}
$C, then the probe volume is sealed and the measurement starts. The
permittivity $\epsilon \left( T\right) $ curves measured two hours after
pressing the powder into disc are shown in Fig. \ref{fig-diel3} for the
sample FAPI-d2. The permittivity had an initially low value $\lesssim 20$,
which sharply raised above 45 around 400~K, the temperature of the $\delta
\rightarrow \alpha $ transformation. Heating was stopped at 430~K in order
to avoid the thermal degradation of the organic component and immediately
followed by cooling at $-1.2$~K/min. During cooling the permittivity
remained high and presented two anomalies at the structural $\alpha
\rightarrow \beta $ and $\beta \rightarrow \gamma $ transitions,
in agreement with the anelastic measurements. There were also an additional
anomaly at 173~K and some structure below the $\beta
/\gamma $ transition, which were reproducible over several temperature
cycles on two different samples.

The $\epsilon \left( T\right) $ curve during heating perfectly reproduces
the cooling, except for the peak/step at $T_{\beta \alpha }$, which is
sharper and higher, but this is probably an effect of incomplete removal of
humidity, as discussed below. Indeed, if some humidity is present during
cooling, it condenses and freezes on the sample and increases the
conductivity and permittivity after becoming liquid during subsequent
heating.

\begin{figure}[tbh]
\includegraphics[width=8.5 cm]{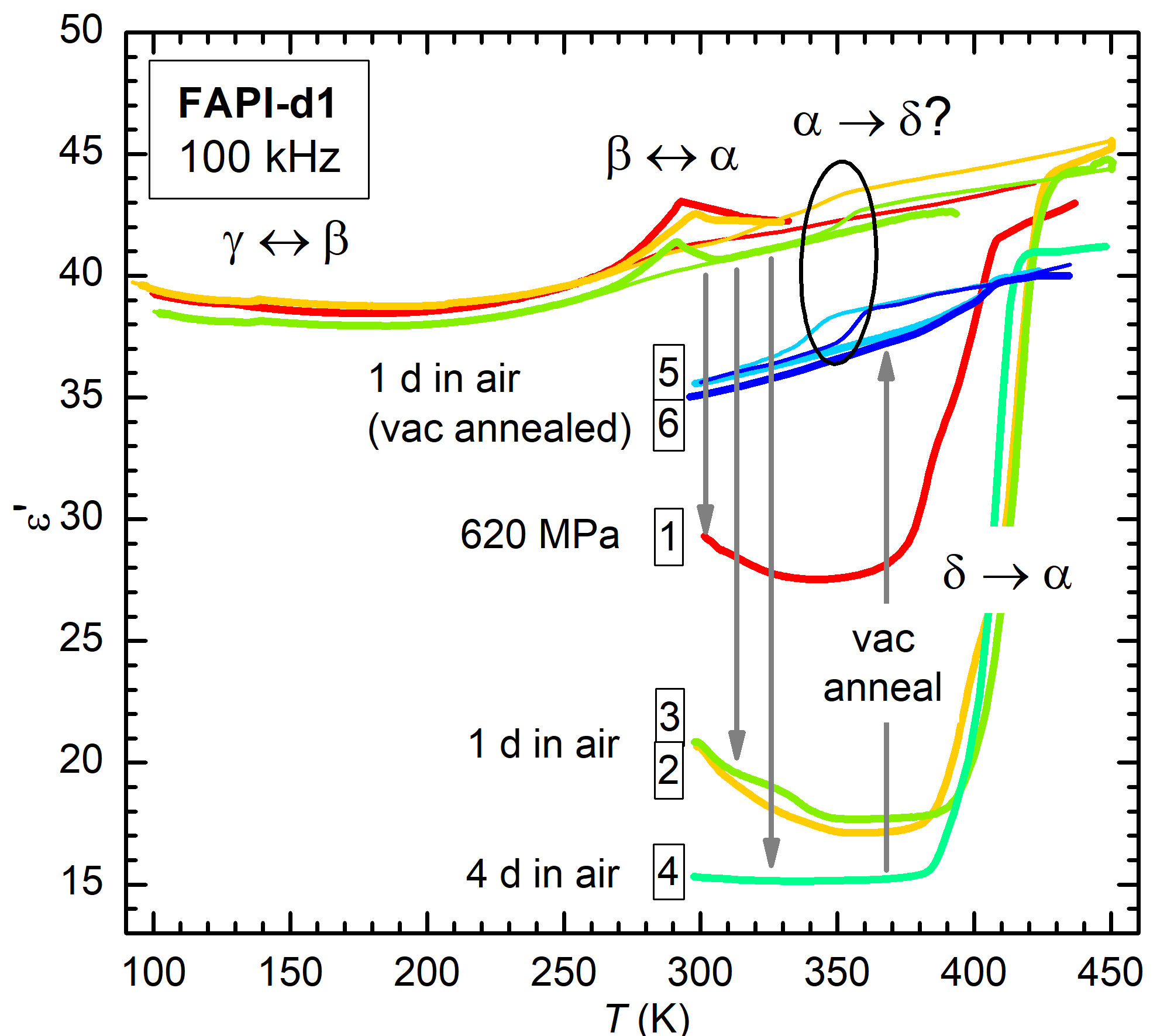}
\caption{Real part of the permittivity of FAPI measured at 100~kHz during
several temperature cycles using the standard procedure for removing
humidity at 40~${{}^\circ}$C. After vacuum annealing, $\protect\epsilon $ is
more stable at its higher level.}
\label{fig-diel1}
\end{figure}

If humidity was removed with the standard procedure described above, after
one day the permittivity dropped again to values even lower than the initial
one of mixed $\alpha $ and $\delta $ phase. This is shown in Fig. \ref{fig-diel1}
reporting results on FAPI-d1. The experiment was repeated twice at one day
interval (curves 2 and 3), yielding the same result, and after four days (4)
the starting permittivity was even lower, until the sample was removed. Its
border free from Ag was still black and was analyzed by XRD (Fig. S3), which indicated
a strongly degraded $\alpha $ phase, without any sign of $\delta $ phase.
Then the sample was annealed for 5~min in a vacuum of $10^{-4}$~mbar at $%
\lesssim 200~{{}^{\circ }}$C and immediately mounted in the dielectric probe
with the standard procedure for flushing out the humidity. This time the
initial permittivity was high (5), and remained so when repeating the
experiment the following day (6). During all these temperature runs there
was a progressive overall decrease of the $\epsilon \left( T\right) $ curves
in the $\alpha ,\beta $ and $\gamma $ sequence of phases, indicating that
some irreversible damage to the sample occurred after each cycle. There is
also a small step when cooling through 350~K, whose amplitude increases at
every cycle; it is tentatively indicated as partial $\alpha \rightarrow
\delta $ transition. These results strongly suggest that
humidity is stored in the sample, cannot be removed by the standard purging
procedure at 40~${{}^{\circ }}$C and is responsible for the degradation of
the $\alpha $ phase, though initially without full conversion to the $\delta
$ phase. Above the $\delta \rightarrow \alpha $ transition temperature, most
of the water is expelled by the sample, but remains trapped into the closed
volume of the probe, so promoting again the degradation of the $\alpha $
phase already during the hours following the measurement. Water was instead
removed by the short treatment at high temperature in vacuum before
measuring the last curves 5 and 6.

\begin{figure}[ht]
\includegraphics[width=8.5 cm]{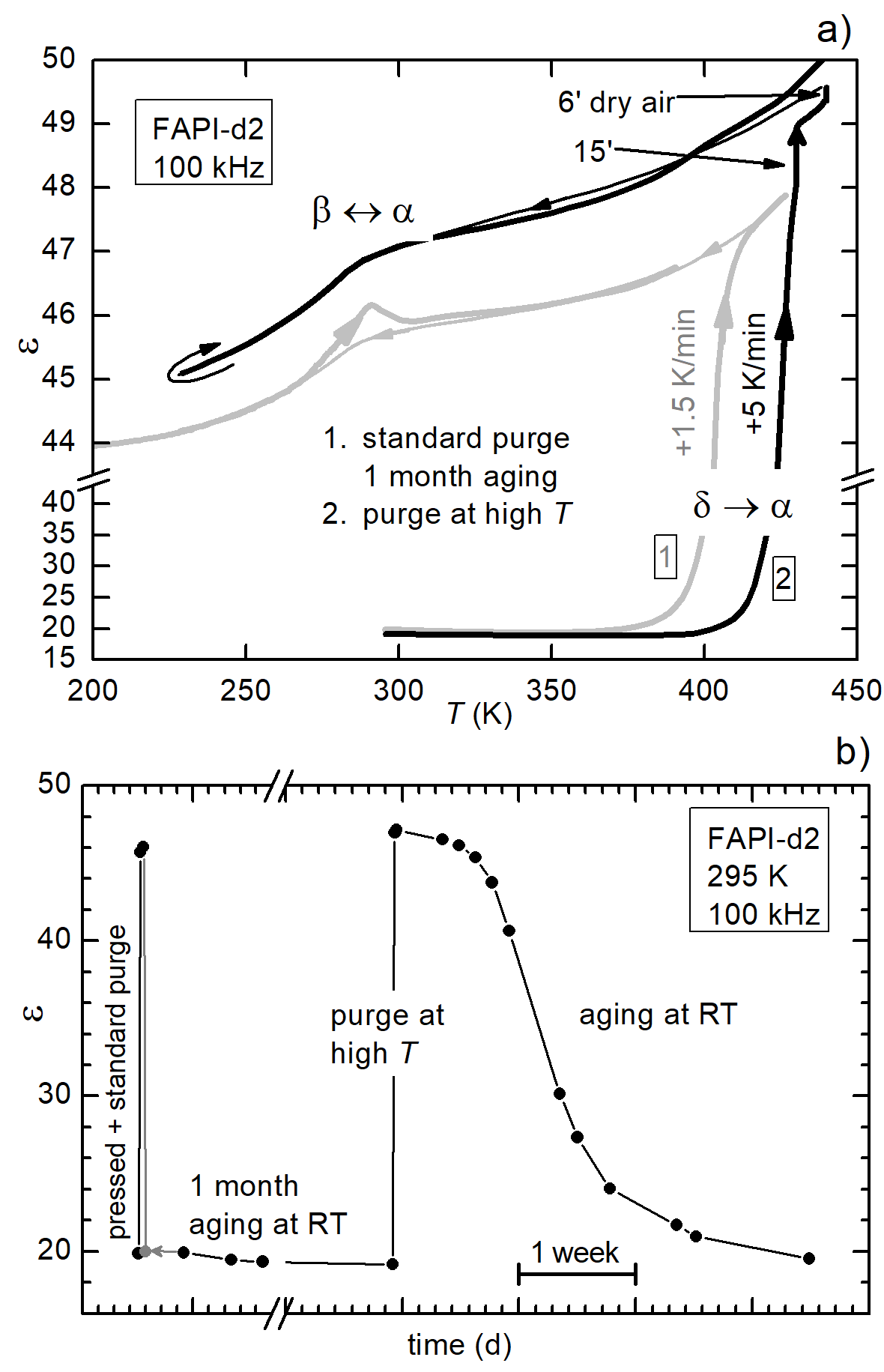}
\caption{Real part of the permittivity of FAPI-d2 measured at 100~kHz during
two temperature cycles and aging. (a) $\protect\epsilon \left( T\right) $
curves measured (1) just after pressing the sample with standard purging from
humidity ar 313~K (same as Fig. \protect\ref{fig-diel3}) and
(2) after aging 1 month and
purging at 440~K. Notice the broken vertical scale. (b) $\protect\epsilon \ $%
measured at 295~K and 100~kHz versus time. }
\label{fig-diel3cycl}
\end{figure}

In order to
confirm this hypothesis, we repeated the measurement of sample FAPI-d2 with
an improved drying procedure.
Figure \ref{fig-diel3cycl}a shows the previous measurement of Fig. \ref%
{fig-diel3} (gray curves 1) with the standard drying procedure followed by
one month aging and the new measurement (black curves 2) with a
protocol where purging is carried out during 6~min at 440~K, above the
$\delta \rightarrow \alpha $ transition (details in the SI).
As expected, the latter procedure was much more effective in
removing water from both the sample and the probe volume: it yielded a more
complete transformation into the $\alpha $ phase, as demonstrated by the
higher level reached by the permittivity, and greatly extended the
stability in time of the $\alpha $ phase, as shown in the lower panel.
As explained in the SI, the degradation kinetics was more than 30 times
slower than previously, and longer dehydration treatment would have
certainly further improved the stability of the $\alpha $ phase.


The experimental results described above can be summarized as follows:

1) According to XRD the loose powder remains stable in the $\alpha $ phase
in normally humid air at room temperature for at least 100 days.

2) Pressing the powder of pure $\alpha -$FAPI into bars or discs at $0.2-0.6$%
~GPa for few minutes induces a partial transformation to the $\delta $
phase and rapid absorption of humidity. The partially degraded phase reverts
to $\alpha $ during the first heating above 400~K, but the
subsequent results on the pressed samples depend on whether the measurements
are carried out in dynamic high vacuum or in a small closed volume.

3) According to the anelastic experiments in high vacuum, the compacted
powder remains stable in the $\alpha $ phase for at least one month,
including one week in air, but after further exposure to air it rapidly
degrades.

4) According to the dielectric experiments in a small closed volume purged
from humidity before each experiment at 313~K, most of the compacted powder
remains stable in the $\alpha $ phase during the full temperature run from
450~K down to 100~K and heating again to above room temperature. However,
already after half day the dielectric permittivity drops back to the
original state of partially degraded $\alpha $ phase, with a seeming
saturation of the degradation process after few days. Yet, repeating the
full temperature cycles reproduces the first experiment, and the whole
process of transformation into $\alpha $ phase and partial degradation can
be repeated several times. The stability of the $\alpha $ phase is enhanced
after purging from humidity is made more effectively at higher temperature.

5) the yellow $\delta $ phase is an advanced stage in the paths to degradation in
the presence of humidity and the dielectric constant and elastic modulus are
strongly sensitive also to the initial and partly reversible stages of
degradation, without visible formation of yellow phases.

At first these observations may seem to add puzzling information to the already
rich known phenomenology, but they can all be easily explained
and harmonized with the existing literature.

The first strong conclusion that can be drawn from 1) and 3) is that $\alpha
-$FAPI, although metastable at room temperature, is much more stable than
previously thought, over a time scale of months and possibly more, depending
on the extrinsic factors that induce the $\alpha \rightarrow \delta $
transformation and/or degradation into PbI$_{2}$ with loss of the organic
part. The nature of these extrinsic factors can be deduced form the above
observations and finds confirmation in the literature.


The discs and bars were pressed at 0.6 and 0.2~GPa, respectively, and the
observation of diffraction peaks of the $\delta $ phase and of yellow grains
is in accordance with the recent report that pressure induces the $\alpha
\rightarrow \delta $ transition at room temperature already at $<0.1$~GPa.%
\cite{JLJ18} On the other hand, there is no agreement with the few other
cases where the effect of pressure on the structure of $\alpha -$FAPI has
been studied: no transition has been found up to 2.5~GPa,\cite{LKG17} or up
to 0.6~GPa in colloidal FAPI nanocrystals\cite{ZCQ18} or transitions to
different phases have been observed,\cite{WGG17,SDW17,ZCQ18}
while nanocrystalline inorganic halide perovskites exhibit better stability.%
\cite{XCQ17,MLL18}
Among all these
data, we find agreement with those or Ref. \citenum{JLJ18} and, regarding
the variety of the available results, we observe that in the small volumes
of the high pressure experiments the issues discussed below related to
adsorbed water or solvents are greatly amplified.


Humidity is perhaps the most acknowledged cause of degradation of the
metal-organic perovskites,\cite{DB16} but there is extreme variability in
the rate and nature of degradation of FAPI, as mentioned in the introduction
and also within the results presented here. The effect of humidity has been
much studied in MAPI, where two different types of degradation have been
recognized, depending whether water condenses at the perovskite surface or
not. Condensed water irreversibly decomposes MAPI into PbI$_{2}$ and HI with
the organic molecule dissolved in water or lost as gas. In addition, the
reaction sequence seems autocatalytic through the degradation products,\cite%
{PBL17} so that beyond a certain point it becomes very fast. On the other
hand, if water does not condense it diffuses in the bulk and forms hydrated
compounds where the network of alkali octahedra is partially broken and
intercalated with the organic and H$_{2}$O molecules;\cite{LHC15} this type
of degradation during a first stage can be reversed by drying the material.%
\cite{DB16}

In FAPI, as in other materials, water condensation should occur
preferentially at grain boundaries. For example in gold, it has been
observed that grain boundaries and terraces on the surface are
preferential sites for water absorption,\cite{GCG01} presumably due to
their favorable topographic conformation. In addition, grain boundaries
often have a higher density of defects, which in the case of the metal-organic
perovskites might favor both the adsorption and reaction of water molecules.
Indeed in FAPI films there are both indirect and direct evidence in this sense.
Voids attributed to the condensation of water during preparation are found to run
along the grain boundaries,\cite{AWA16} and the degradation in FAPI films
exposed to humidity has been observed via Kelvin probe
force microscopy to initiate at grain boundaries.\cite{YKY18} In addition,
combined scanning TEM and energy loss spectroscopy detected the presence of
OH$^-$ ions at the FAPI grain boundaries, a sign of reacted water.\cite{AWH16}


With these premises, the above observations $1)-4)$ become clear.
The superior stability, even in humid air, of $\alpha -$FAPI in loose
powder with respect to the compacted form is explained with the fact that
the powder has few grain boundaries, if any, whereas in the compacted
powder, as in films, each grain is circumscribed by grain boundaries acting
as water condensation centers.

Both types of degradation under humidity described above, reversible
intercalation of H$_{2}$O and subsequent decomposition, considerably lower
the elastic moduli and dielectric constant, because they break the network
of Pb-I bonds and eliminate the polyhedral spaces between octahedra, where
the organic molecules are free to reorient. It is therefore impossible to
distinguish among the various stages of hydration and occurrence of partial $%
\alpha \rightarrow \delta $ transformation from sets of curves like those of
Figs. $1-3$. The positive step in $\epsilon $ and $M$ when heating through
400~K is labeled $\alpha \rightarrow \delta $ because it corresponds to the
known temperature of that transition, but probably it includes the
contribution from dehydration of the material. Analogously, the smaller step
sometimes observed during cooling in $\epsilon \left( T\right) $ is labeled $%
\ $"$\alpha \rightarrow \delta ?$ ", but it must also include rehydration
from the water remained trapped in the closed volume of the dielectric
probe. The hydration and subsequent reactions are frozen during cooling to
lower temperatures, but proceed when back to room temperature, where $%
\epsilon $ drops to less than half within the next day. The proof that water
drives the degradation process is given by the prolonged stability of the $\alpha $
phase during aging at room temperature (curve 2 in Fig. \ref{fig-diel3cycl}a
and Fig. \ref{fig-diel3cycl}b) after the improved dehydration of the sample
and elimination of water from the closed volume.
A similar result is obtained by a preventive
drastic dehydration at $10^{-4}$~~mbar and 200~${{}^\circ}$C (curves 5,6 of
Fig. \ref{fig-diel1}).
Moreover, the bar for the anelastic measurements remains stable in air for
9 days, after a thorough dehydration in high vacuum for few days.
This is in accordance with results in films, where a thermal treatment
after preparation is found to improve the stability.\cite{YKY18}

Considering that the TGA runs do not show measurable losses of water on $%
\alpha -$FAPI powder, we must conclude that water is mainly adsorbed after
pressing the powder, due to the formation of the grain boundaries and
probably even more to a higher hygroscopicity of the $\delta $ phase, and is
released only at high enough temperature, probably $\geq 400$~K. Therefore,
if the outgassing of the sample is performed near room temperature or not at
all, the adsorbed water remains in the sample and starts to be released only
near or above the $\delta \rightarrow \alpha $ transition and, if
trapped in a closed volume, it is readsorbed during cooling and promotes
degradation and possibly the reverse $\alpha \rightarrow \delta $
transformation. This is what happened in our first dielectric runs (curves
1-4 of Fig. \ref{fig-diel1}).

The conclusion drawn from this discussion is that, although $\alpha -$FAPI
is metastable at room temperature, having $\delta -$FAPI a lower
energy, the energy barrier for converting the $\alpha $ into the $\delta $
phase is so high that no measurable conversion
occurs at room temperature for at least 100 days (results from XRD on powder).
The $\alpha \rightarrow \delta $ reaction is probably catalyzed by water,
together with other
degradation pathways, which include hydration and decomposition; the latter
has been found to occur even spontaneously in vacuum at the surface of MAPI.%
\cite{DAP15}


As pointed out in the Introduction, there is no consistency among the
various reports on the stability of $\alpha $-FAPI in the literature. This
may be due to several factors, such as different preparation protcols,
different types of samples, testing conditions and methods. Yet, it seems
that the major discrepancy between the present observations and the
literature is with the careful investigation of FAPI powder by neutron
diffraction in Ref. \citenum{CFP16}. In particular, in that experiment it is found that
a large quantity ($10-12$~g) of FAPI powder undergoes both the $\delta
\rightarrow \alpha $ transition at 350~K during heating and the $\alpha
\rightarrow \delta $ at 290~K during cooling in a closed volume filled with
He, whereas we do not observe the latter neither in vacuum nor in air, when
starting from pure $\alpha $-FAPI. Instead, similarly to other neutron and
X-ray diffraction experiments,\cite{FSL16,WGG18} the octahedral tilt
transitions of the perovskite structure are observed during cooling.
Considering the well controlled conditions and effectiveness of the
experimental technique of Chen \textit{et al.},\cite{CFP16} the discrepancy
calls for an explanation, which we think is in the similarity between the
conditions of that experiment and of our dielectric measurements with incomplete
purging of humidity. In fact, in both cases the starting phase is completely
or partially $\delta $, most likely hydrated, so that the water expelled
by the sample at high temperature remains trapped in the experimental volume
and drives the partial reversion to the $\delta $ phase during cooling.

The picture presented here is also in agreement with the previous observation%
\cite{FSL16} that the powder of $\alpha $-FAPI obtained after heating $%
\delta $-FAPI overnight in a vacuum oven at 120~${{}^{\circ }}$C was stable
for $20-30$ days, whereas heating $\delta $-FAPI in an evacuated ampoule
under static vacuum yields a black powder that reverts to yellow after few
minutes. In the latter case the considerable amount of water contained in $%
\delta $-FAPI remains completely trapped in the ampoule and drives the
reverse $\alpha \rightarrow \delta $ reaction immediately after cooling to room
temperature.

It would also be interesting to reconsider the experiments where the stability is
found to be highly enhanced in nanocrystalline FAPI. The observations
include stable photoluminescence of colloidal nanocrystals with protected
surface,\cite{LOT17} where the stabilization of the cubic phase is explained
as a general effect of the small size of the particles, which enhances the
surface-to-volume ratio.\cite{LOT17,ZCQ18} Similarly, the observations of
stability over several months of FAPI nanostructures protected by long chain
alkyl or aromatic ammonium cations has been explained in terms of change of
balance between bulk and surface energy after the surface functionalization,%
\cite{FWW17} while the stability of $\alpha -$FAPI embedded in alumina
nanotubes has been explained with the fact that the latter limit the
possibility of expansion of $\alpha -$FAPI during degradation.\cite{GZK18}
In view of the stability of our micrometric loose powder and thoroughly
dried compacted powder, we suggest that also the stability of nanostructured
FAPI might alternatively be explained in terms of protection of $\alpha -$%
FAPI from humidity by the surface treatments or by the nanotubes and in
terms of reduced fraction of grain boundaries in these single nanocrystals.


Our observations show that, even if the $\alpha $ phase of FAPI is
metastable with respect to the yellow $\delta $ phase, its kinetic trapping
at room temperature lasts for much longer than previously thought. The
present data set the period of perfect stability of $\alpha -$FAPI to at
least 100 days, corresponding to the duration of our tests, but this time
range exceeds seven months if previous experiments on FAPI embedded in
alumina nanotubes\cite{GZK18} or with functionalized surfaces\cite{FWW17}
are reinterpreted in terms of protection against humidity. This stability
also includes the absence of spontaneous decomposition.

It appears that by far the most important factor inducing degradation of $%
\alpha -$FAPI is humidity at grain boundaries, and that the procedures used
for dehydrating the material are often insufficient. Since grain boundaries
are also present in films, these findings should help in devising strategies
for improving the stability of $\alpha -$FAPI without modifying its
composition, particularly by improving the initial dehydration process and
subsequent surface protection.
This is especially true in the cases of intentional exposure to moisture
during preparation in order to improve the grain morphology through
solvent annealing.\cite{PHM16,CAV18}

\begin{acknowledgement}

The authors thank Paolo Massimiliano Latino (ISM-Tor Vergata) for
his technical assistance, Sara Notarantonio (ISM-Montelibretti) for
the assistance in the synthesis, Marco Guaragno (ISM-Tor Vergata)
for his technical support in X-ray experiments,
Dr. Cristina Riccucci for the DTA-DSC measurements and Dr. Gloria Zanotti
for fruitful discussions.

\end{acknowledgement}

\begin{suppinfo}

The following files are available free of charge.
\begin{itemize}
  \item SuppInfo.pdf:
 Synthesis, Thermal Analysis of $\alpha- $ and $\delta- $FAPI; Pressing;
 X-ray diffraction, particle size, degraded samples; Complex Young's
 modulus method and data of degraded sample; Dielectric measurements.
\end{itemize}

\end{suppinfo}

\bibliographystyle{plain}
\bibliography{refs}

\end{document}